\documentclass[conference]{IEEEtran}
\IEEEoverridecommandlockouts
\usepackage{graphicx}
\usepackage{amsmath}
\usepackage{amssymb}
\usepackage{booktabs}
\usepackage{algorithm}
\usepackage{textcomp}
\usepackage{stfloats}
\usepackage{url}
\usepackage{verbatim}
\usepackage{graphicx}
\usepackage{cite}
\usepackage{caption}
\usepackage{booktabs}
\usepackage{algorithmic}
\usepackage{algorithm}
\usepackage{array}
\graphicspath{ {imgs/} } 

\def\BibTeX{{\rm B\kern-.05em{\sc i\kern-.025em b}\kern-.08em
    T\kern-.1667em\lower.7ex\hbox{E}\kern-.125emX}}
\begin{document}

\title{A Novel Score-CAM based Denoiser for Spectrographic Signature Extraction without Ground Truth}

\author{\IEEEauthorblockN{Noel Elias}
\IEEEauthorblockA{\textit{University of Texas at Austin, USA} \\
nelias@utexas.edu}
}

\maketitle

\begin{abstract}
Sonar based audio classification techniques are a growing area of research in the field of underwater acoustics. Usually, underwater noise picked up by passive sonar transducers contains all types of signals that travel through the ocean and is transformed into spectrographic images. As a result, the corresponding spectrograms intended to display the temporal-frequency data of a certain object often include the tonal regions of abundant extraneous noise that can effectively interfere with a ‘contact'. So, a majority of spectrographic samples extracted from underwater audio signals are rendered unusable due to their clutter and lack the required indistinguishability between different objects. With limited clean true data for supervised training, creating classification models for these audio signals is severely bottlenecked.

This paper derives several new techniques to combat this problem by developing a novel Score-CAM based denoiser to extract an object's signature from noisy spectrographic data without being given any ground truth data. In particular, this paper proposes a novel generative adversarial network architecture for learning and producing spectrographic training data in similar distributions to low-feature spectrogram inputs. In addition, this paper also a generalizable class activation mapping based denoiser for different distributions of acoustic data, even real-world data distributions. Utilizing these novel architectures and proposed denoising techniques, these experiments demonstrate state-of-the-art noise reduction accuracy and improved classification accuracy than current audio classification standards. As such, this approach has applications not only to audio data but for countless data distributions used all around the world for machine learning.
\end{abstract}


%%%%%%%%% BODY TEXT
\section{Introduction}
The ongoing development of the Internet as well as the advancement of multimedia technologies have allowed for the increase of the dissipation, utilization, documentation, and creation of digital audio signals. With many different audio signals being circulated through both public and private sectors of society, the demand for tools to analyze these signals is also on the rise. In addition, with the increased utilization of big data analytics, the need for machine learning techniques to analyze these signals for signal classification, detection, and prediction are heavily sought after.  	

One such field where such audio classification techniques are being heavily researched is underwater acoustics.

Within the ocean, there are a variety of sound pulses from objects including submarines, ships, divers, and even different animals. With the evolution of underwater technology, the growing field of underwater acoustics has allowed parties to analyze sound pulses using a variety of invasive and non-invasive techniques. In particular, the increase in sonar has allowed individuals to use sonar-based devices that are much more powerful than current radio signals being used for acoustic analysis.

Sonar, which is based on echolocation, works by propagating sound energy into the water and recording the reflected energy for analysis. Sonar utilizes microphones known as hydrophones which can pick up the mechanical noise in the water and convert them to electrical energy utilizing a transducer. These sonar signals can either be passive or active. Active Sonar transducers emit an acoustic signal or pulse of sound into the water. If an object is in the path of the sound pulse, the sound bounces off the object and returns an “echo” to the sonar transducer. If the transducer is equipped with the ability to receive signals, it measures the strength of the signal. Passive Sonar systems are used primarily to detect noise from marine objects such as submarines or ships and marine animals like whales. Unlike active sonar, passive sonar does not emit its own signal, it only detects sound waves coming toward it.  In this proposal, we will mainly discuss the signals produced by passive sonar that listens to sound made by other vessels. 

Sonar audio signals are then often converted into digital audio signals where they are transformed into frequency domain-based data utilizing Fourier transformations. These frequency representations of a signal are then transmuted into a variety of graphical representations to illustrate the interconnecting domain relationships of the signal. One such visualization includes spectrograms which display the temporal frequency sampling of the signal. 

\begin{equation}
A_k = \sum_{n=0}^{N-1} W_N^{kn} a_n
\label{dft.eq}
\end{equation}
where
$$
W_N = e^{-i {2 \pi \over N}}
$$

To determine the FFT of a discrete signal as shown in Equation 1, A[k] is computed concurrently on odd and even subsequences of a signal that reduces the problem to O(NlogN) complexity \cite {maklin_2019}. 

These spectrograms provide the necessary visual feature embeddings that many machine learning algorithms like Convolutional Neural Networks (CNNs), Support Vector Machines (SVMs), regression models, etc. can use for multi-class classification of different objects of audio signals. So, utilizing the following methodologies many underwater audio signals collected through sonar-based microphone devices can be sampled, detected, and classified.

\section{Problem}
Most CNN-based machine learning models require sufficient amounts of true data for each class so that a model’s weights can learn differences between deeper features of different classes.  As a result, training image data for most supervised models must be clear, correctly labeled, and diverse to create efficient and accurate machine-learning models. This implicit requirement within CNN-based approaches poses a problem for most types of underwater audio signal analysis. 

Sonar picks up all types of signals that travel through the ocean. This includes intended objects but is not limited to other human activity, marine animals, echoes, etc. As a result, spectrograms intended to display the temporal-frequency data of a certain object often include the tonal regions of abundant extraneous noise that can effectively interfere with a ‘contact’. So, a majority of spectrographic samples extracted from underwater audio signals are rendered unusable due to their clutter and lack of the required indistinguishability between different objects. With limited clean true data for supervised training, creating classification models for audio signals is severely bottlenecked.

In addition, this problem is compounded by the lack of methods to properly denoise spectrographic samples and extract the necessary nonlinear tonal regions of different objects within the data. As a result, even if an object’s tonal regions are known beforehand, limitations exist in the ability to automatically extract/identify these regions from spectrograms for future training and identification purposes. 

Thus, while massive amounts of sonar data can be collected, they are not able to be processed or analyzed in any significant automated manner. This proposal proposes an approach to classify and extract signatures of different objects within spectrographic regions given limited ground truth data. 

\section{Literature Review}
The problem of image denoising and signature extraction particularly for spectrographic representation of audio signals is a relatively new problem statement with few tapped research outputs. However, the larger problem of image denoising has been greatly researched within the machine learning community. 

The current state-of-the-art method of image denoising revolves around utilizing filtering, particularly non-linear filtering techniques to remove Gaussian distributed image noise. A drawback with this technique however is that these spatial filters eliminate noise at the cost of image blurring which loses the sharp features of the data or in this spectrographic case, the identifying features \cite{fan2019brief}. Another method of image denoising requires image transformation that often comes in the form of component analysis and deep denoiser based on Convolutional Neural Networks (CNNs). 

CNN denoisers try to learn a mapping function that transforms the input noisy images to denoised images by optimizing a loss function on noisy-clean training pairs. One of the most common and popular methodologies for this process include auto-encoders and denoiser CNNs (DnCNNs) in which these feed-forward deep networks performed much better in being able to achieve denoising effects on the data. The problem with this approach revolves around cases where there is a lack of fundamental truth data or denoised images for a CNN model to be trained on. As such, because of the limited training data, most CNN-based denoisers are rendered unusable. 

To generate additional data to be used for image denoising, various approaches have been used by researchers to learn and replicate noise distribution within data. Soltanayev et al. showed that using Stein’s unbiased risk estimator (SURE), researchers were able to learn the noise distributions and simulate additional training data to use for classifying the images\cite{soltanayev2018training}. In addition, utilizing a generative adversarial network model, researchers were also able to generate and learn noise distributions for improved training data diversity when utilizing a DnCNN denoiser\cite{tran2020gan}. In regards to spectrogram denoising, the few proposals of image denoising revolving around pixel-wise semantic image segmentation to remove all extraneous classes of the target class to result in clean images \cite{mallawaarachchi2008spectrogram}. This approach is once again limited to the massive amount of truth data.  

Thus, in cases where there are little to no truth data for clean spectrographic signatures, particularly spectrograms, there is a lack of research understanding and use of viable alternatives. This proposal demonstrates a deep Score-CAM-based approach to derive signatures of noisy images with limited to no truth data at all.

\section{Approaches}
Essentially, we want to first identify spectrogram signals that belong to the same class and create an identification model for this class. Then, we want to denoise the input spectrograms by extracting the tonal regions specific to the target class from the spectrogram. 

\begin{figure}[h]
\centering
\includegraphics[width=\linewidth]{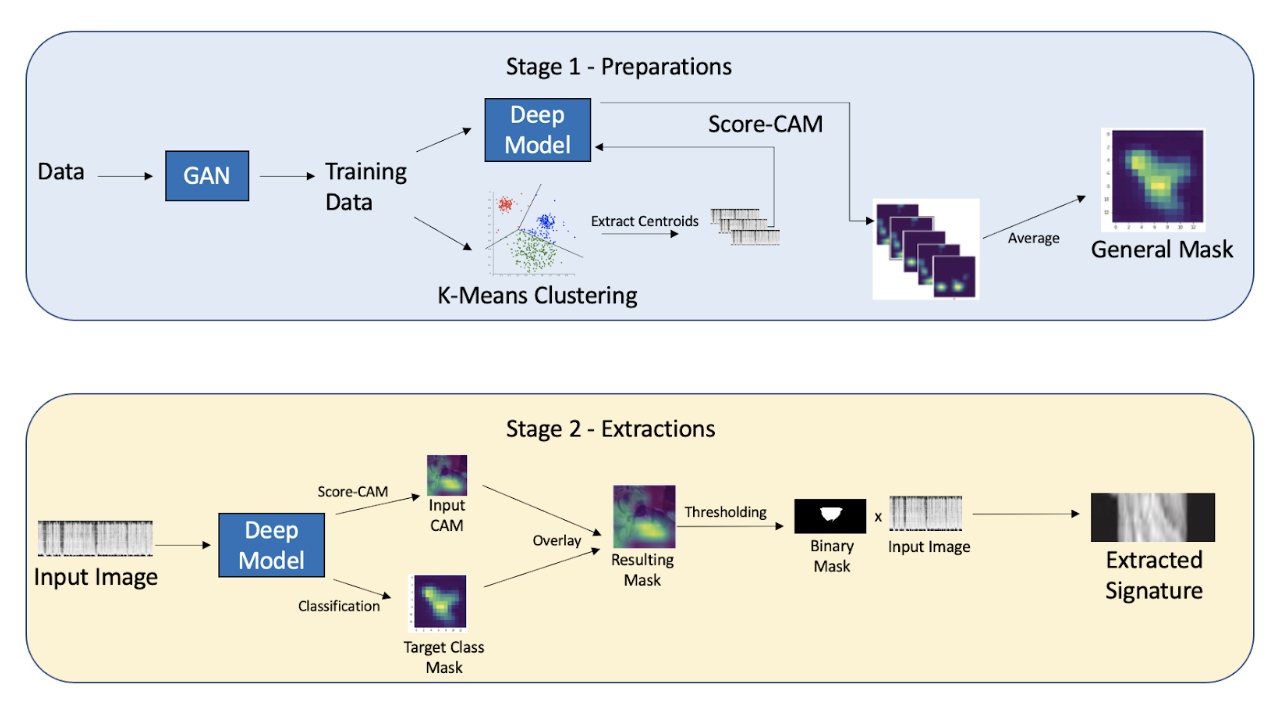}
\caption{Overall workflow of signature extraction.}
\label{fig:workflow}
\end{figure}

The general approach to this problem involves a variety of steps as shown in Figure \ref{fig:workflow}. First, using a Generative Adversarial Network (GAN), additional data is created in a similar distribution so that a deep neural network model can be efficiently trained to classify the signal.

After a classification model is trained on the data, a class activation mask is generated that encompasses the location of the tonal regions for that target class using a clustering algorithm as well as Score-CAM. For an input image, the generated class mask is overlaid with the image CAM mask and used to extract regions of interest for the target class. The resulting image is the denoised spectrogram. 

\subsection{Data Generation}
To combat the lack of limited spectrographic training ground truth data, this proposal proposes the utilization of Generative Adversarial Networks (GANs). GANs are modeled to utilize instances of labeled spectrograms to generate new spectrographic data containing new possible tonal regions alongside different noise distributions to represent additional training data. GANs utilize a generative and discriminative module during training in which the generator is back-propagated to generate high-probability instances of output images that are characteristically similar to existing truth data as determined by the discriminator.

The GAN training occurs in two steps, first, the discriminator model is trained to accurately classify between the real data set and a static generator. This loss function emphasizes the difference between the predicted vs true class of the input images of the generator and existing data samples. Then, utilizing a uniform distribution of noise as random input for our generator, we generate discriminator classifications alongside the training data and backpropagate to use the gradient to only change generator weights.  

To maintain the static nature of both modules while training the other, an alternating training sequence was also used where the discriminator and generative model were each trained for several epochs before switching to the other. This continued until the convergence of the discriminator performance as its accuracy decreased to about 50\% indicating truly random classification accuracy based on real and generated images. The customized loss function for both generator and discriminator training was the Wasserstein loss function \cite{arjovsky2017wasserstein}. 

The Wasserstein loss function as implemented by \cite{tensorflowcontributors}, is derived from the earth mover distance formula determining the distance between two probabilistic distributions by giving a true metric. In essence, similar to a triplet loss function, the discriminator or ‘critic’ simply outputs a larger number for contrasting images while smaller numbers for real images. Thus, the one-line discriminator loss minimizes D(x) – D(G(y)) where D is the discriminator output of the input image and G is the output of the generator for some random seed-like value y. The one-line generator loss thus simply minimizes the discriminator output D(G(y)) to get smaller values.  The Wasserstein GAN loss algorithm is described in Algorithm 1. In addition, as opposed to minimax loss for GANs, Wasserstein loss can remedy the vanish gradient problem by enforcing local minima by retracting stabilization points. 

\begin{figure}[h]
\centering
\includegraphics[width=\linewidth]{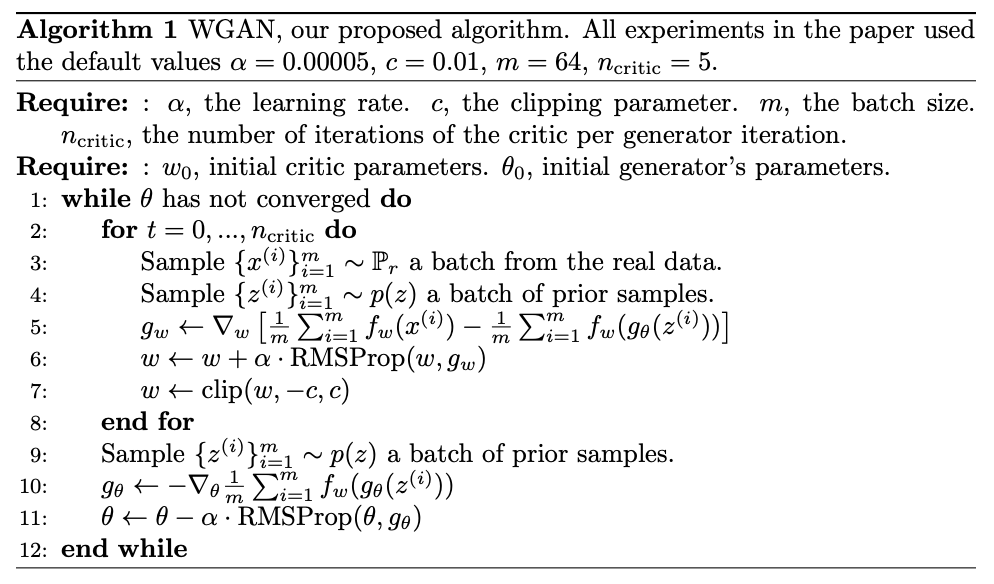}
\end{figure}

\begin{figure}[h]
\centering
\includegraphics[width=\linewidth]{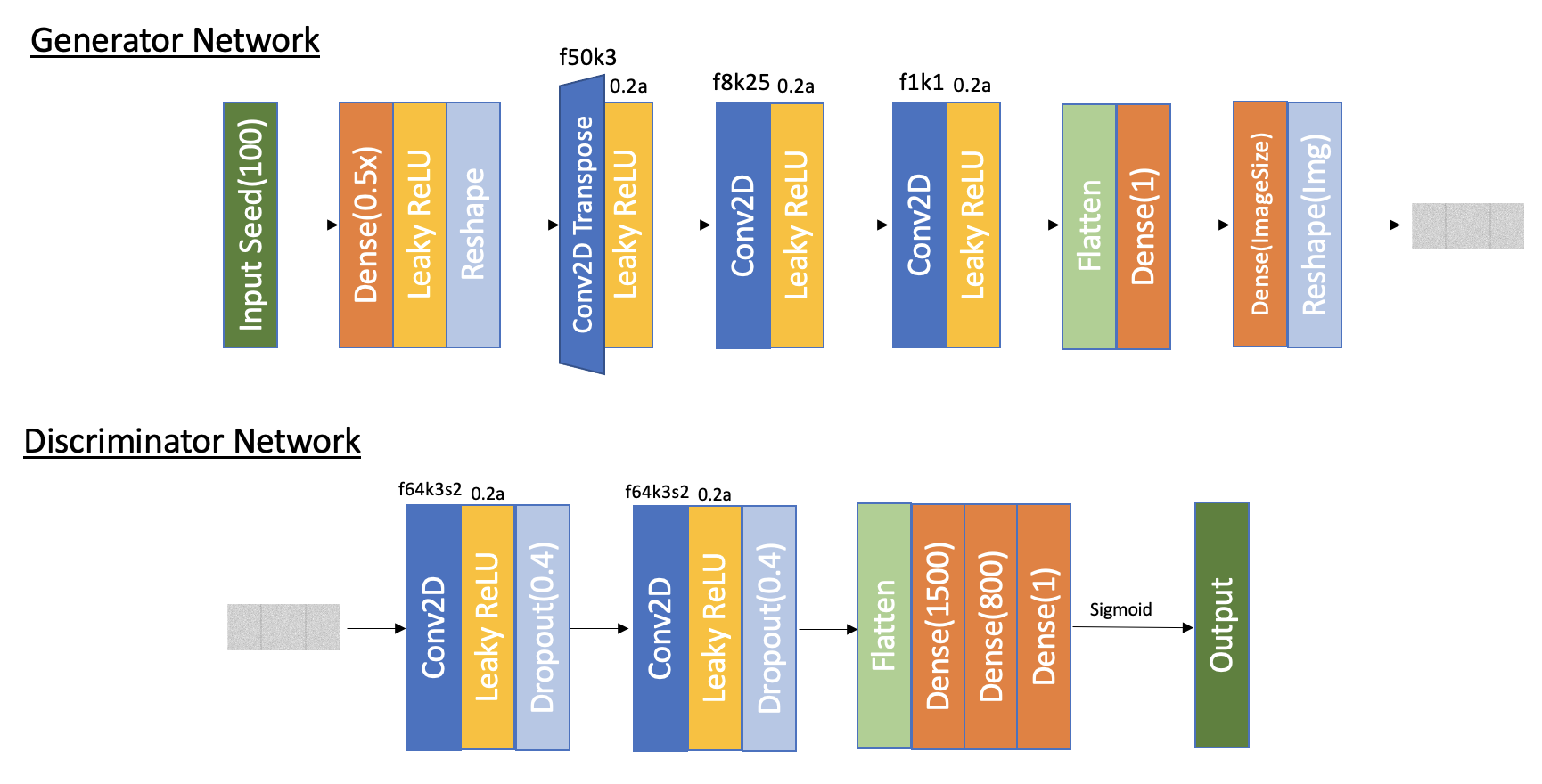}
\caption{Utilized GAN architecture.}
\label{fig:ganarch}
\end{figure}

Based on the similar distribution of output images that we wanted, different methodologies of the types of GANs were experimented on to provide the most efficient and accurate modeling. The architecture utilized is illustrated in Figure \ref{fig:ganarch}. This was based on previous research in creating deep models utilizing low-feature spectrograms. In addition, the specific parameters of the Wasserstein GAN, or WGAN in this case were input latent dimension of 100 seed, training for 100 epochs, with a batch size of around 20 for around 100 samples of data. In addition, binary cross-entropy loss was used on both generator and discriminator networks. So, with the additionally generated data, deep learning multi-class models are trained to identify the signals with the new ample data.

\subsection{Image Clustering}
To create the most accurate mask containing the tonal regions of the target class, this proposal suggests using the class activation maps (CAMs) of the most representative samples of a class to determine the overall saliency map of the target class.  This proposal proposes utilizing the K-Means clustering algorithms to classify all the images within a data set into different clusters and extract the centroids of each group to find the most representative images of all the different data distributions within that class. 

To further enhance this clustering algorithm, a K-Means++  initialization for centroids was utilized \cite {arthur2006k}. Instead of randomly picking points within the data set, to be the initial centroids, K-Means++ suggests only randomly picking a single point as a centroid and maximizing the distance to the other point as the next centroid as shown by Algorithm 2. In essence, each centroid is the point farthest from the current centroid until k centroids have been selected. As a result, the algorithm runs in O(logN) time. In addition, to determine the optimal number of clusters needed for each data set, the elbow method was used to determine the point of diminishing returns for k-clusters. 

\begin{algorithm}
    \caption*{\textbf{Algorithm 2} K-means clustering algorithm}
        Choose primary centroids $v_{k}$
        . Then, compute the membership degree of all feature vectors in all clusters
        \begin{equation*}
            u_{ki}  = \frac{1}{ \sum_{j=1}^C ( \frac{D^{2}(x_{i} - v_{k})}{D^{2}(x_{i} - v_{j})})^\frac{2} 
            {m-1}}   
        \end{equation*}
\end{algorithm}

\begin{figure*}[h]
    \centering
    \includegraphics[width=\textwidth]{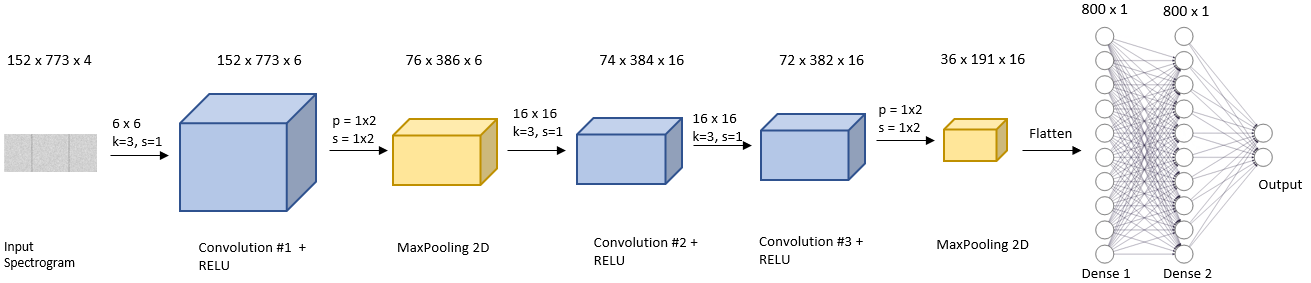}
    \caption{This illustrates Spec-CNN architecture used to train classification models.} 
    \label{fig:specCNN}
\end{figure*}

The input data for this KMeans clustering was not the raw images themselves but the feature embeddings of the dense layer of the classification model used to classify the normalized image into its potential predicted class. The classification model architecture (Spec-CNN) that was chosen utilized previously researched techniques when working with spectrographic data and is shown in Figure \ref{fig:specCNN}. This model was trained with all the different available data of different classes. Then an input image was inputted into the model to find its predicted class. Once, the target class was found, the image and the other members of its class were extracted in the form of embeddings of the second dense layer of the SpecCNN architecture and used for the KMeans clustering algorithm. The embeddings were utilized over the actual images themselves to preserve feature weightage and compress the image data for better clustering results.  

The KMeans clustering model utilized all the embeddings of known images of that target class to cluster the data. Once this model was finished clustering the data, the centroids were extracted from the data set.

It was postulated that the most representative samples would be better left as images within the dataset themselves rather than a centroid’s vector decoded into a potential image. As such, for each centroid, the nearest correlated embedding of an image in that cluster was identified by minimizing the Euclidean distance or L2 distance formula between the image points. The corresponding spectrogram images for each cluster were then utilized as the most representative images of the target class data and were then passed on to generate general masks. 

The specific training parameters of the KMeans clustering model included a random state of 0, a max iteration of 300, and k-means++ centroid initialization formula, and the KElbowVisualizer library to determine the elbow point for the most optimal centroid number. 

\subsection{ScoreCAM}
Now, to denoise the spectrograms we first need to identify the parts of a spectrogram that correspond to the tonal regions of a specific signal. To do this we can use Score-Weighted Visual Explanations for Convolutional Neural Networks (Score-CAM). 

After training a Convolutional Neural Network (CNN) on the known spectrograms of the known signals of a certain ship, class activation mappings (CAM) can be utilized on an input spectrogram of that class to understand regions of the spectrogram that most correlate to the class activation within the mode. Using this mask, these regions can be extracted from the spectrogram and stored as denoised spectrograms representing the plain temporal-frequency data of an identified signal. 

Unlike previous iterations of CAM techniques including Gradient Based CAM (GradCAM), Score-CAM does not rely on back-propagating of the dense layer of a CNN to create a rough localization of the class activation \cite{selvaraju2017grad}, but instead utilizes a perturbation-based approach. This lack of gradient-based CAMs indicates less disturbance within the saliency maps of the image due to the random noise caused by gradients. 

\begin{figure}[h]
\centering
\includegraphics[width=\linewidth]{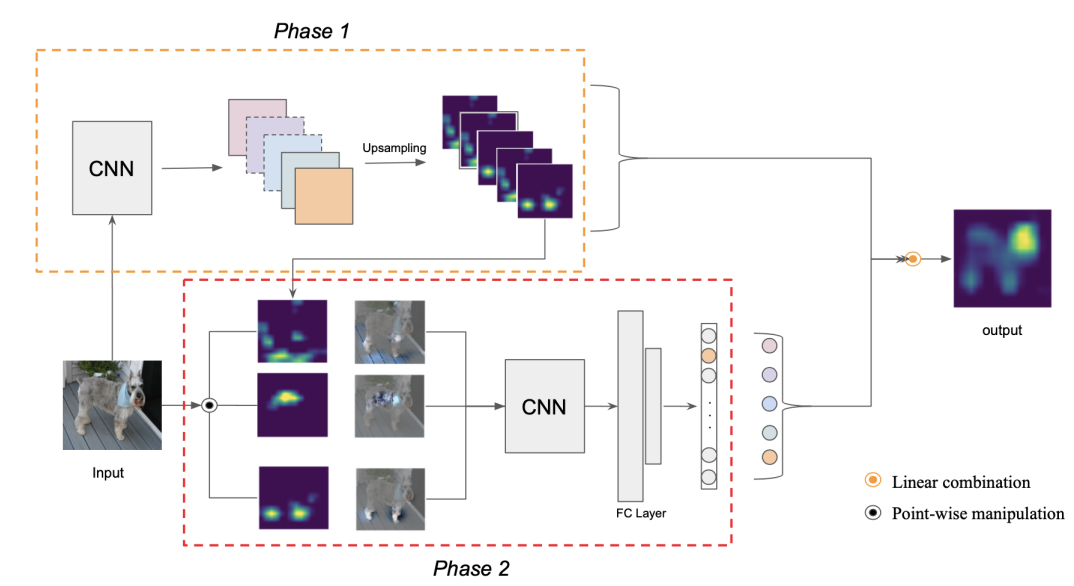}
\caption{Sample Score-CAM steps.}
\label{fig:scorecam}
\end{figure}

In Score-CAM \cite{wang2020score}, a forward pass is run on an input image where the activations are extracted from the last convolutional layer of the model. These activation maps are then unsampled to fit the input image size and normalize the values of the layer. The Hadamard product of the input image as well as the normalized activation map is then fed into a SoftMax layer to get the scores of all the classes, especially the target class. Lastly, the ReLU activation function is applied to the linear combination of the target class score and the sum of all the activation maps to result in a final activation map representing the saliency mask of the class on the image \cite{mishra_2021}. This is shown in Figure \ref{fig:scorecam}. Equation \ref{dft.eq} illustrates the Score-CAM formula where C is the Channel-wise Increase of Confidence score for an activation map A \cite{wang2020score}.

\begin{equation}
L_{Score-CAM}^{c} = ReLU(\sum_{k} a_k^{c} A_l^{k})
\label{dft.eq}
\end{equation}
where
$$
a_k^{c} = C(A_l^{k})
$$

The resulting mask of the input image can then be overlaid on the image to determine the parts of the image that are activating the target class. Based on the masking, the noise can be removed from the image based on instances where the image is less highlighted by the mask. This results in a denoised image of the signal with the corresponding tonal regions. 

\subsection{Mask Generation}
As mentioned, after the clustering algorithm was run on the data sets, the centroids of the clustering model were extracted. These centroids served as the most representative images of the data set and were imputed into Score-CAM. The saliency maps for each of these centroid images were simply averaged together to result in the overall mask of the target class. 

This average general mask is overlaid with the Score-CAM mappings of that image itself when run through the CNN. This is done to highlight any tonal regions within the spectrographic image that might be outliers found by the model but not present in the centroid-based mask. So, the mask for any image is a combination of a general mask for the target class as well as the image-specific mask. 

\subsection{Signature Extraction}
Lastly, to finally extract the class-specific tonal regions from the spectrogram, the generated mask is applied to the image to only extract the class’s tonal regions rather than any extraneous interference. 

To do this and provide an instance of generality for other masks and input images, the proposal proposes using a region of interest (ROI) approach using the saliency map on the image. In this process, the compounded mask undergoes thresholding to convert the saliency map into a binary mask of black and white. This thresholding was also experimented on by determining the specificity of the model between the ranges of 0-1. The mask will contain an array of points of either 1 or 0. Points with the value 1 will belong to the target class and thus will be kept while points marked as 0 will are identified as noise. This mask can then be multiplied against the input image to remove all extraneous noise from the sample to reveal the model’s signature. The resulting 0 points on the signature can be scaled to 255 to output white space indicating a lack of tones in that frequency and at that time. Thus, the signature is extracted from the input image using this denoising technique. 

\section{Results}
\subsection{Data Generation}
Utilizing the WGAN architecture mentioned above, the WGAN model involving the discriminator and the generator models was trained and then extracted to be able to generate new data samples from random seed values. 

\begin{figure}[h]
\centering
\includegraphics[width=\linewidth]{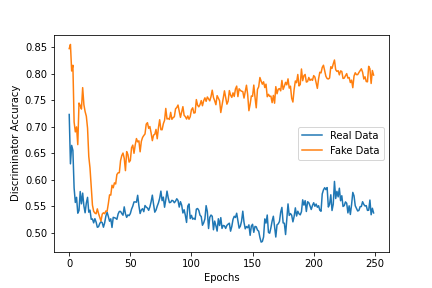}
\caption{WGAN generator and discriminator acuracy}
\label{fig:wgan_acc}
\end{figure}

The training accuracy of the WGAN model can be seen in Figure \ref{fig:wgan_acc}. The optimal model weights for the generator lie at around epoch 150 where the discriminator is only able to distinguish the real data from the fake data with about a 50\% accuracy meaning that both samples are similar enough that we get a truly random classification. 

\begin{figure}[h]
\centering
\includegraphics[width=\linewidth]{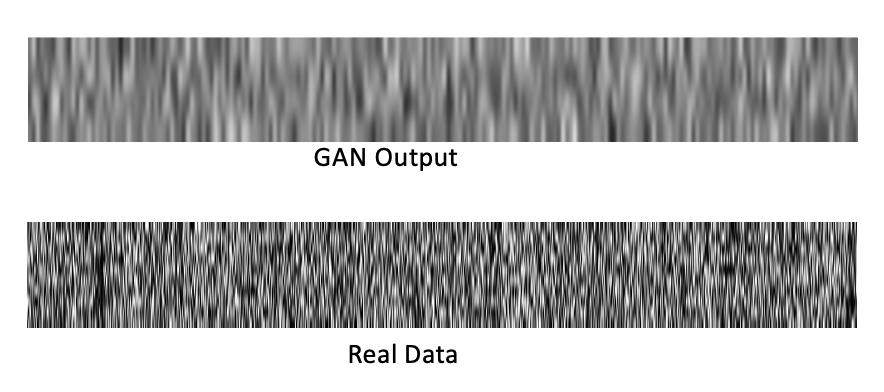}
\caption{WGAN Output vs Real Data output}
\label{fig:wgan_data}
\end{figure}

Lastly, a sample prediction was compared to the general training data for the target class. Ultimately, the generated WGAN data represented possible spectrographic images in a truly random data set. In essence, the generated spectrograms could have been valid spectrograms given a certain noise distribution. However, with the low-feature limitation of the target spectrograms in general, it was impossible to determine if the generated data was actually representative of the target class. The small feature differences between the WGAN data and the existing target dataset could indicate completely different object tones. It is also very difficult to check if all unique identifying features of a target class were included in a WGAN output as shown in Figure \ref{fig:wgan_data}. 

So, to be safe, the generated WGAN data was only partially used for training a model to classify different spectrographic samples and not used for denoising methods of individual images as the signature extraction could be contaminated by inaccurate data. 

However, the WGAN itself can solve the original problem of the lack of training data to train classification CNN-based models of audio spectrograms of sonar noise by improving the accuracy/diversity of the data set by a distinguishable amount. 

\subsection{Image Clustering}
\begin{figure}[h]
\centering
\includegraphics[width=\linewidth]{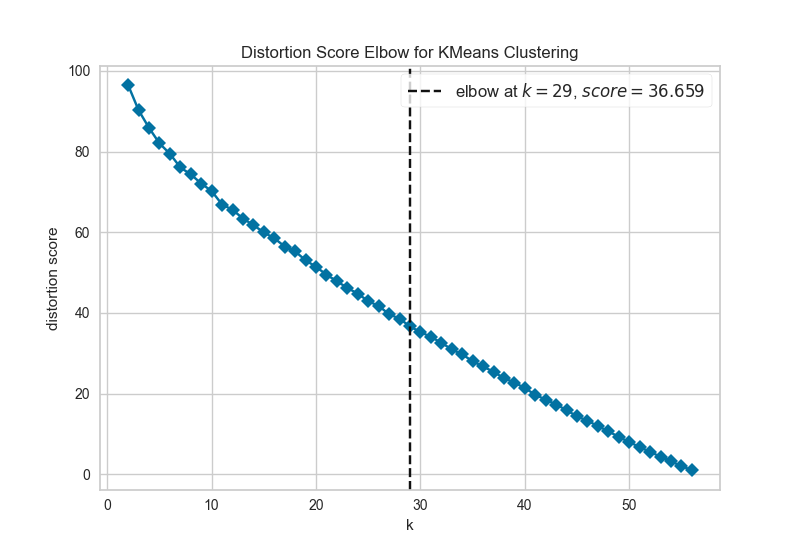}
\caption{KMeans Clustering Elbow Point}
\label{fig:elbow_point}
\end{figure}

Within the image clustering, the first step was to figure out the optimal amount of centroids to be used by finding the elbow point of the inertia metric with all possible centroid values as shown in Figure \ref{fig:elbow_point} where the optimal number of clusters was 29 with an inertia score of ~36. The seemingly lack of cluster elbow was much more visible when analyzing the points graphically or with other clustering methodologies including density based clustering. 

Using this value, the KMeans algorithm was run on the normalized image's embeddings to determine the clusters of different data features. The resulting clustering was 2D and thus included a lot of point overlap due to the normalized nature of the embeddings. Afterward, the points closest to the centroids were extracted and utilized to determine the necessary masks for the input image. 

\subsection{Mask Generation}
For generating the general mask of that class, the Score-CAM arrays of the most representative images of each class as calculated above were selected and averaged together. 

Afterward, the input image was also run through Score-CAM to generate its own class activation map for the predicted class of that image. This array was overlayed onto the general mask to generate the image-specific mask that can be used for signature extraction.

\begin{figure}[h]
\centering
\includegraphics[width=\linewidth]{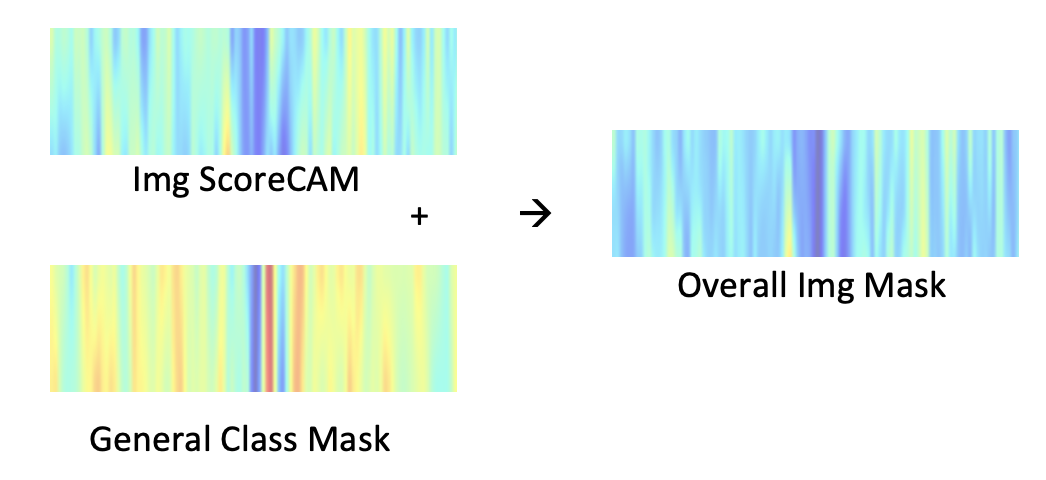}
\caption{Image Specific Mask for Input Image}
\label{fig:imgmask_gen}
\end{figure}

\subsection{Signature Extraction}
The signature can now be extracted from the input image by applying the image-specific mask to the image. This was done by first determining a specificity threshold to convert regions of interest with a certainty score of x and above as the class-specific tones. In this case, this threshold was determined to be 0.65. 

\begin{figure}[h]
\centering
\includegraphics[width=0.8\linewidth]{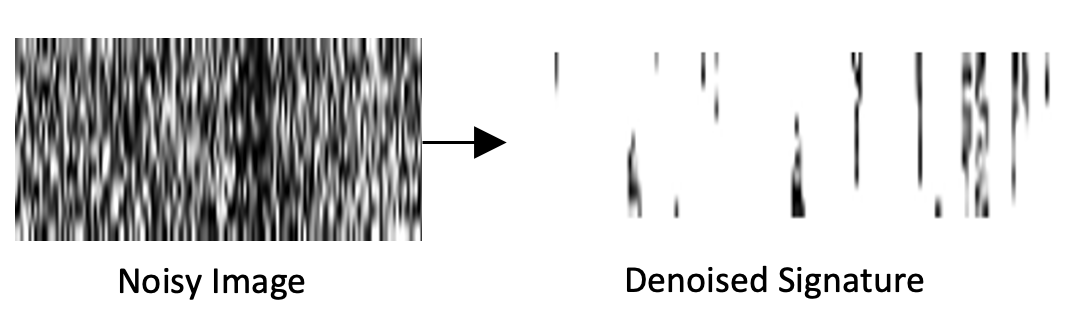}
\caption{Denoised and extracted signature of image according to prediction class.}
\label{fig:signature_extract}
\end{figure}

The threshold of 0.65 was applied to the image-specific mask to create the binary mask for extraction. Then this binary mask was applied to the image to activate and extract our image's true signature for the predicted class. The resulting image is properly denoised to show the tonal regions of the specific object it contains and illustrates the most basic and unique features of that class as shown in Figure \ref{fig:signature_extract}. 

\section{Experiments}
Several experiments in regards to this general approach as well as comparisons to main-stream standards were tested in this paper. 

The first experiment involved using auto-encoders as denoisers on sample images to extract the noise and extract a clean signature. This was ultimately eliminated because it was a supervised method requiring ample amounts of truth-extracted signature data. In many cases, researchers might not even have access to the truth data involving a denoised sample. They might be relying on the natural denoising architecture of CNNs to almost drown the background noise of audio spectrograms to get to the most primary tonal regions of an object's sound and thus image features. So, this auto-encoder approach would not be a viable alternative denoiser for extracting spectrographic signatures. 

Next, the proposed denoiser was tested against itself using different parameters/approaches to determine the most optimal approach and hyper-parameters for a signature-extracting denoiser. 

The tested approaches in these experiments include the proposed denoiser, as well as the denoiser with different confidence/specificity thresholds, an auto-encoder-based mask generator, and a reverse signature extractor. 

The different confidence/specificity thresholds that were tested in the signature extraction included 0.65, 0.75, and 0.85 intervals. These resulting signatures were compared with others. 

\begin{figure}[h]
\centering
\includegraphics[width=\linewidth]{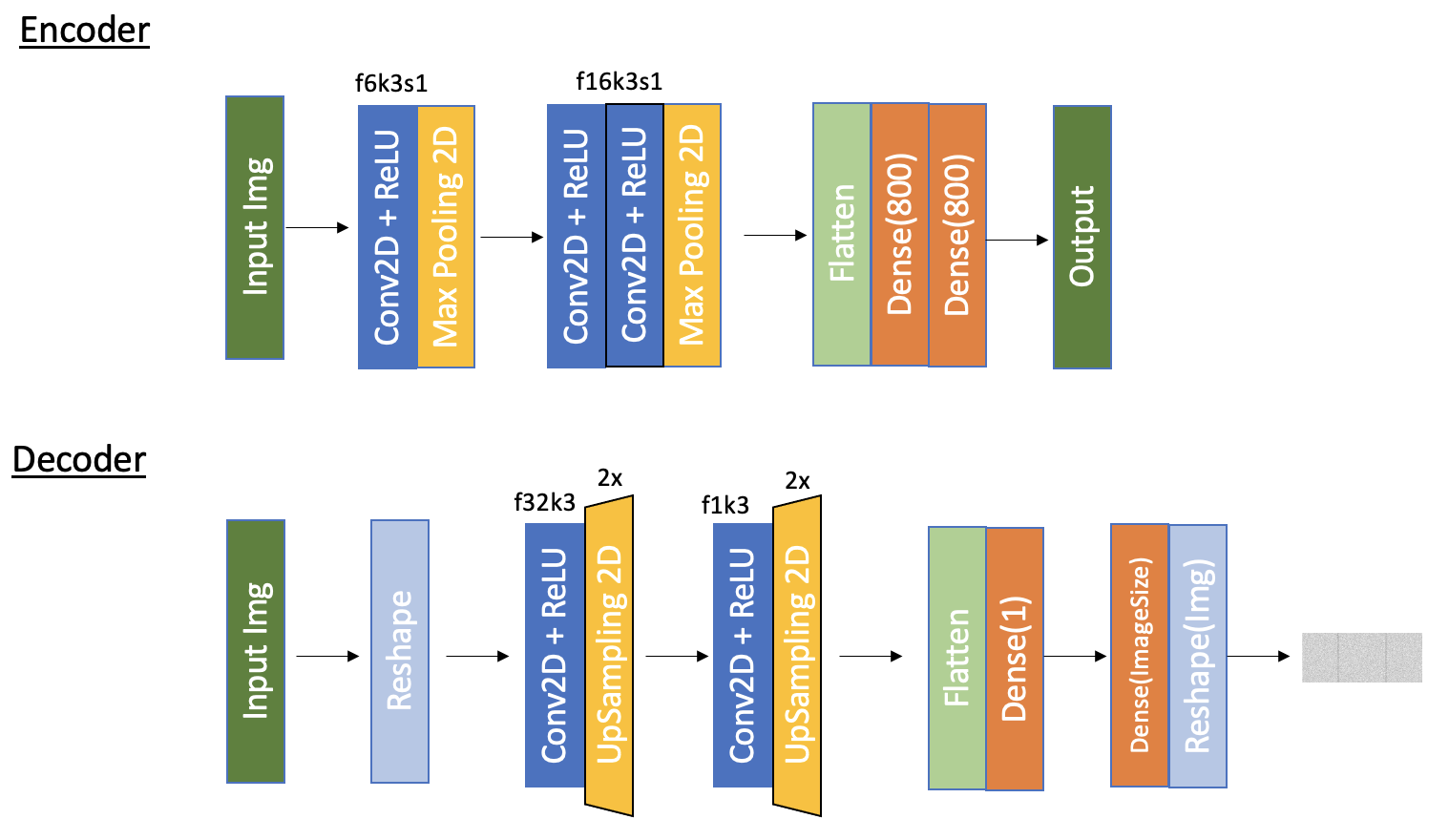}
\caption{Auto-encoder to convert centroid embeddings to spectrogram images}
\label{fig:autoencoder}
\end{figure}

The auto-encoder-based mask generation approach involves extracting the centroids from the clustering model and then using these centroids as the most representative samples of the target class to then combine and form a general mask. However, to convert these centroids into the form of possible embeddings and transform them into a possible spectrogram, an auto-encoder was utilized. The encoder structure of the auto-encoder was the Spec-CNN model architecture where the output was the second dense layer before the final classification nodes. The decoder architecture was similar to the WGAN generator model and is shown in Figure \ref{fig:autoencoder}. The auto-encoder was trained for 200 epochs with a batch size of around 40 utilizing a binary cross-entropy loss function. 

In addition to this method, a reverse approach to the proposed signature extraction method was tested. In this method, the input image was inputted and classified by the trained classification model. Then, once the target class has been identified, the proposed denoising process is applied to all other classes of that model except for the predicted class. Once all these masks are generated, they are applied to the image the corresponding regions are removed from the image. As a result, all other noises and possible object tones are removed from the image leaving us only with the target class's tonal regions. Thus, this method removes and extracts the noise from the input rather than the proposed approach of extracting the predicted class's regions and leaving the noise behind. So, in one method we extract the signature and in the other, we extract the noise and leave the signature.

\begin{figure}[h]
\centering
\includegraphics[width=0.8\linewidth]{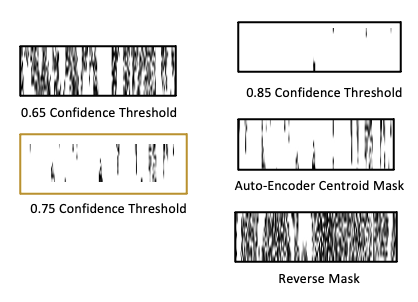}
\caption{Resulting extracted signatures of experiment}
\label{fig:experiment}
\end{figure}

\begin{table}[ht]
\caption{Experiment results}
\begin{tabular}[t]{l>{\raggedright}p{0.2\linewidth}>{\raggedright\arraybackslash}p{0.2\linewidth}>{\raggedright\arraybackslash}p{0.2\linewidth}}
\toprule
&Removed Noise & Overwritten Tones & Intersecting Tonal Regions\\
\midrule
0.65 CT & 26\% & 4\% & 15+\\\
\textbf{0.75 CT} &86\% & 8\% & 2\\
0.85 CT &95\% & 59\% & 0\\
AE Centroids  &79\% & 7\% & 5\\
Reverse CT &14\%  & 3\% & 15+\\

\bottomrule
\end{tabular}
\end{table}%

The experiment measured and analyzed the resulting signature extraction of the different approaches based on overall appearance regulated by previous knowledge of correct signatures, percent of general noise removed, false negative tonal regions, and tonal regions that appeared on other class signatures indicating they were not the class's unique features. 

Based on the quantitative and qualitative results from these different experiments illustrated in Figure \ref{fig:experiment} and Table 1, the most optimal denoising approach was the proposed approach with a 0.65 confidence threshold. The most optimal results were approached with a high degree of removed noise without removing necessary tonal information (low false negatives) and efficiently distinguishing the input image's features from other classes (low intersecting tones) The 0.75 threshold experiment indicates around 86\% of wanted noise removed, only 8\% false negative regions and only 2 intersection tonal regions with other classes indicating that most of the signature is unique. 

The 0.65 threshold resulted in a too-dense image mask and the 0.85 threshold resulted in a scattered mask that drowned out most features. The 0.65 threshold barely removed noise and as a result, left many tonal regions of other classes within the signature. In addition, while the 0.85 threshold removed most of the noise, it also removed the tonal regions of the class. The reverse mask also resulted in a much more dense signature as noise that is usually removed during a class Score-CAM analysis was not a primary region of any of the other target classes the model was trained on. Because we do not have all possible classes to classify all pixels of data, some pixels may not be removed as they have not been recognized, causing additional noise to be left in the signature. This resulted in low removal of noise causing a lot of tonal overall.  Lastly, the auto-encoder centroid mask worked the second most efficiently in our experiments with similar results to the 0.75 threshold given 79\% noise removal and fewer tonal region overlaps. However, because the general mask was created by possible centroid embeddings and not rooted in actual data, this method was not preferred. 

So, overall the proposed approach with a threshold of around 0.7-0.75 seems to extract the most amount of unwanted noise from an image compared to other methods. 

\section{Key Takeaways}
Here are some key decisions and experimental advances that were made:
\begin{itemize}
    \item When creating an image-specific mask, in some cases it might be better to simply add the general class mask as well as the image-specific mask as some images might be outliers compared to the general data distribution. So, adding in instead of maximizing the pixel map values will retain general mask features. 
    \item When clustering embeddings and training models in general for WGANS, auto-encoders, classification, etc it was best to utilize fewer convolutions and layers of multi-node dense layers to help the mode differentiate between features in a vector form. 
    \item Utilizing Score-CAM for identifying ROI within the image is itself a noise remover due to the score-based CAM approach. As a result, averaging multiple masks will essentially cancel out possible noises that might not be covered in all spectrograms. 
\end{itemize}

\section{Conclusion}
This paper has presented a  novel strategy to tackle the ongoing challenge of denoising audio-based data, particularly spectrograms for machine learning classification and recognition using very little ground truth data. 

Namely, this paper has demonstrated that noisy audio signals often in a  spatial-temporal form can be transformed to clean spectrographic signatures that can be used for machine learning models to accurately learn and classify audio signals. This paper has also most importantly developed a Score-CAM-based denoising workflow that can be efficiently trained on various distributions of data to remove extraneous noise from input images without needing previously extracted signatures as ground truth. 

Specifically, by first generating additional data, creating a classification model, clustering class data, and creating a Score-CAM-based image mask we can extract the tonal regions of a class from a noisy input. First, using a Wasserstein-GAN, a generator is created to supply a classification model with sufficient data to efficiently discriminate and classify between different classes. Next, the training data of the predicted class is clustered to determine the most representative images near each centroid. Afterward, these images are averaged into a general Score-CAM mask alongside the input image Score-CAM result and are thresholded to mask the input image and extract the cleaned image signature. 

This semi-supervised approach proved to be the most efficient and accurate as opposed to current deep denoising methods including auto-encoders and is the first of its kind due to its lack of independence of training data. Specifically, we found that the proposed approach with 	$\approx$ 0.75 thresholding confidence gives state-of-the-art performance and denoising results. 

Overall, we hope that this paper marks the start of utilizing this general deep Score-CAM-based denoising technique to be able to not only accurately classify and extract spectrographic signatures but any noisy image one might encounter without the need for truth data. 

{\small
\bibliographystyle{ieee_fullname}
\bibliography{egbib}

\begin{thebibliography}{10}\itemsep=-1pt

\bibitem{arjovsky2017wasserstein}
Martin Arjovsky, Soumith Chintala, and L{\'e}on Bottou.
\newblock Wasserstein generative adversarial networks.
\newblock In {\em International conference on machine learning}, pages 214--223. PMLR, 2017.

\bibitem{arthur2006k}
David Arthur and Sergei Vassilvitskii.
\newblock k-means++: The advantages of careful seeding.
\newblock Technical report, Stanford, 2006.

\bibitem{tensorflowcontributors}
Tensorflow Contributors.
\newblock Tensorflow gans.

\bibitem{fan2019brief}
Linwei Fan, Fan Zhang, Hui Fan, and Caiming Zhang.
\newblock Brief review of image denoising techniques.
\newblock {\em Visual Computing for Industry, Biomedicine, and Art}, 2(1):1--12, 2019.

\bibitem{maklin_2019}
Cory Maklin.
\newblock Fast fourier transform, Dec 2019.

\bibitem{mallawaarachchi2008spectrogram}
Asitha Mallawaarachchi, SH Ong, Mandar Chitre, and Elizabeth Taylor.
\newblock Spectrogram denoising and automated extraction of the fundamental frequency variation of dolphin whistles.
\newblock {\em The Journal of the Acoustical Society of America}, 124(2):1159--1170, 2008.

\bibitem{mishra_2021}
Divyanshu Mishra.
\newblock Demystifying convolutional neural networks using scorecam, Jul 2021.

\bibitem{selvaraju2017grad}
Ramprasaath~R Selvaraju, Michael Cogswell, Abhishek Das, Ramakrishna Vedantam, Devi Parikh, and Dhruv Batra.
\newblock Grad-cam: Visual explanations from deep networks via gradient-based localization.
\newblock In {\em Proceedings of the IEEE international conference on computer vision}, pages 618--626, 2017.

\bibitem{soltanayev2018training}
Shakarim Soltanayev and Se~Young Chun.
\newblock Training deep learning based denoisers without ground truth data.
\newblock {\em Advances in neural information processing systems}, 31, 2018.

\bibitem{tran2020gan}
Linh~Duy Tran, Son~Minh Nguyen, and Masayuki Arai.
\newblock Gan-based noise model for denoising real images.
\newblock In {\em Proceedings of the Asian Conference on Computer Vision}, 2020.

\bibitem{wang2020score}
Haofan Wang, Zifan Wang, Mengnan Du, Fan Yang, Zijian Zhang, Sirui Ding, Piotr Mardziel, and Xia Hu.
\newblock Score-cam: Score-weighted visual explanations for convolutional neural networks.
\newblock In {\em Proceedings of the IEEE/CVF conference on computer vision and pattern recognition workshops}, pages 24--25, 2020.

\end{thebibliography}
}

\end{document}